\documentclass[review]{elsarticle}

\usepackage{lineno,hyperref}
\usepackage{pifont}
\usepackage{amsmath,amssymb,amsfonts}
\usepackage[normalem]{ulem}
\usepackage{graphicx}
\usepackage{textcomp}
\usepackage{multirow}
\usepackage{longtable}
\usepackage{makecell}
\usepackage{lscape}
\usepackage{slashbox}
\usepackage{nicefrac, xfrac}
\usepackage{bbm}
\usepackage{booktabs}
\usepackage{caption}
\usepackage{makecell}
\usepackage{boldline}
\usepackage{subcaption}
\usepackage{booktabs}
\usepackage{afterpage}
\usepackage{setspace}
\usepackage[table]{xcolor}
\usepackage{hhline}
\usepackage[final]{pdfpages}
\usepackage[linesnumbered,ruled,vlined]{algorithm2e}
\SetKwInput{KwInput}{Input}                
\SetKwInput{KwOutput}{Output}              
\SetKwInOut{KwData}{Procedure}

\usepackage{tikz}
\usetikzlibrary{tikzmark}

\usepackage{algpseudocode}
\usepackage[linesnumbered,ruled,vlined]{algorithm2e}
\algnewcommand\algorithmicforeach{\textbf{for each}}
\algdef{S}[FOR]{ForEach}[1]{\algorithmicforeach\ #1\ \algorithmicdo}

\usepackage[font=small,labelfont=bf]{caption}

\usepackage[skip=0.2\baselineskip]{caption}
\usepackage{array}
\newcolumntype{L}[1]{>{\raggedright\arraybackslash}p{#1}}
\newcolumntype{C}[1]{>{\centering\arraybackslash}p{#1}}
\newcolumntype{R}[1]{>{\raggedleft\arraybackslash}p{#1}}

\begin{document}
\begin{frontmatter}
\title{Semi-Supervised Learning with Online Knowledge Distillation for Skin Lesion Classification}

\author{Siyamalan Manivannan\corref{mycorrespondingauthor}}
\cortext[mycorrespondingauthor]{Corresponding author. Email: \href{mailto:siyam@univ.jfn.ac.lk}{siyam@univ.jfn.ac.lk}}
\address{Department of Computer Science, Faculty of Science, University of Jaffna, Sri Lanka.}	




\begin{abstract}	Deep Learning has emerged as a promising approach for skin lesion analysis. However, existing methods mostly rely on fully supervised learning, requiring extensive labeled data, which is challenging and costly to obtain. To alleviate this annotation burden, this study introduces a novel semi-supervised deep learning approach that integrates ensemble learning with online knowledge distillation for enhanced skin lesion classification.
Our methodology involves training an ensemble of convolutional neural network models, using online knowledge distillation to transfer insights from the ensemble to its members. This process aims to enhance the performance of each model within the ensemble, thereby elevating the overall performance of the ensemble itself. Post-training, any individual model within the ensemble can be deployed at test time, as each member is trained to deliver comparable performance to the ensemble.  This is particularly beneficial in resource-constrained environments. Experimental results demonstrate that the knowledge-distilled individual model performs better than independently trained models.
Our approach demonstrates superior performance on both the \emph{International Skin Imaging Collaboration} 2018 and 2019 public benchmark datasets, surpassing current state-of-the-art results. By leveraging ensemble learning and online knowledge distillation, our method reduces the need for extensive labeled data while providing a more resource-efficient solution for skin lesion classification in real-world scenarios.
\end{abstract}

\end{frontmatter}

\section{Introduction}
Skin cancer, particularly Melanoma, ranks as the most prevalent cancer globally. Early detection significantly boosts the five year survival rate of Melanoma to over 90\%~\cite{ge2017skin}. Hence, prompt and accurate identification of skin lesions holds paramount importance. Traditionally, dermatologists rely on visual inspection and manual evaluation of these lesions, a process susceptible to subjectivity and time-intensive analysis. Deep learning techniques, on the other hand, offer a transformative solution to automate and improve the precision of skin lesion classification, ensuring timely and objective predictions.

Deep learning models proposed for skin lesion classification heavily rely on fully supervised learning (FSL)\cite{ge2017skin,zhuang2018skin,codella2017deep,yu2016automated,zhang2019attention}, necessitating a substantial volume of labeled data for training. However, acquiring and annotating such datasets is laborious and expensive. To address this challenge, semi-supervised deep learning (SSL) approaches have gained attention in skin lesion analysis~\cite{peng2023faxmatch,zhou2023refixmatch,zhou2023fixmatch,Mahmood2024,liu2020semi}. These methods utilize both labeled and unlabeled data during training, enhancing model performance even with limited labeled data compared to FSL.

\emph{Ensembling}~\cite{ganaie2022ensemble} is a widely adopted technique in deep learning aimed at addressing the limitations of individual models, combating overfitting, and improving overall accuracy and robustness, and has been widely used for skin lesion analysis under FSL setting~\cite{Shahin,codella2017deep,zhuang2018skin}. It involves training multiple models independently and combining their predictions to get the final prediction. While ensembling offers enhanced performance compared to single models, it comes with significant computational expenses and demands large memory resources, even during test time. To alleviate these burdens, \emph{Knowledge Distillation} (KD)~\cite{hinton2015distilling} is employed, which aims to maintain the performance of the ensemble model while reducing computational requirements and memory usage, rendering it suitable for deployment in resource-constrained environments. In ensemble KD~\cite{asif2019ensemble}, the learned knowledge from the ensemble (referred to as the \emph{teacher}) is transferred to a smaller \emph{student} model to emulate the teacher's behavior. This is accomplished by transferring the knowledge embedded in the teacher model to the student model, typically through soft target labels.

Based on training strategies, KD can be categorized into two main approaches: offline~\cite{hinton2015distilling} and online~\cite{guo2020online,borza2023teacher}.
In offline KD, a two-step process is followed: the ensemble model or teacher is initially trained, followed by training the student model while keeping the teacher fixed. However, in this approach, the teacher receives no direct benefit from KD.
Conversely, in online KD~\cite{zhu2018knowledge,guo2020online}, multiple student models (comprising the ensemble) are trained simultaneously, with their combined predictions serving as the teacher's predictions. This collective knowledge is distilled into each individual model (student) within the ensemble, thereby reducing the computational complexity associated with the two-step training process. Moreover, empirical evidence suggests that online KD often leads to higher accuracy gains compared to offline KD~\cite{wang2021knowledge}, benefiting both the student and the teacher.

While ensembling~\cite{Shahin,codella2017deep,zhuang2018skin} and KD~\cite{khan2022knowledge} techniques have been applied in the literature for skin lesion analysis within the FSL setting, their utilization in the SSL setting, especially in the domain of skin lesion analysis, remains relatively unexplored.
This work proposes a novel SSL technique for skin lesion classification, integrating ensemble learning and online KD. The approach involves training a set of Convolutional Neural Network (CNN) models simultaneously. Online KD is employed to disseminate the collective knowledge acquired from the ensemble model to each individual model within the ensemble. This process enhances the classification performance of individual models, thereby elevating the overall performance of the ensemble. During testing, any knowledge-distilled individual model from the ensemble, especially in resource-constrained environments, or the entire ensemble can be utilized. Experimental results on the publicly available benchmark datasets, ISIC 2018 and ISIC 2019, demonstrate superior performance compared to single independently trained models, establishing the proposed approach as the new state-of-the-art.

The key contributions of this work are as follows:
\begin{itemize}
	\item Introduction of an ensemble-based SSL approach, integrating online KD for skin lesion classification.
	\item Demonstration  of enhanced performance of knowledge-distilled individual models over independently trained counterparts, thereby reducing computational and memory burdens during testing.
	\item Achievement of new state-of-the-art results on ISIC 2018 and ISIC 2019 skin lesion analysis benchmark datasets, validating the efficiency and efficacy of the proposed methodology.
\end{itemize}

The subsequent sections are structured as follows: Section~\ref{sec:rw} presents a discussion on related work. Section~\ref{sec:method} elaborates on the methodology. Section~\ref{sec:exp} reports the experiments along with discussions. Lastly, Section~\ref{sec:con} provides a summary of the conclusions.

\section{Related Work\label{sec:rw}}
Early efforts in skin lesion analysis relied on hand-crafted features and machine learning classifiers. For instance, Ganster et al.~\cite{ganster2001automated} utilized a set of features, including shape, with a KNN classifier for melanoma recognition. However, with the advent of deep learning, several approaches have emerged that surpass traditional hand-crafted feature-based methods for skin lesion classification. Therefore, this section delves into recent advancements in deep learning, both FSL and SSL approaches, for skin lesion classification.

\subsection{Fully Supervised Learning based approaches}
The majority of research in skin lesion analysis relies on FSL approaches. Various deep learning architectures, including ResNet~\cite{yu2016automated,zhuang2018skin}, DenseNet~\cite{hassan2020skin}, SENet~\cite{zhuang2018skin}, have been explored for this purpose. In~\cite{zhang2019attention}, an attention-based residual learning was proposed, which places greater emphasis on lesion regions than normal skin regions, resulting in improved performance. Ensemble learning has also been employed to enhance the classification performance. For instance, Zhuang et al.\cite{zhuang2018skin} trained different CNN models, and aggregated their results for final predictions. Similarly, in~\cite{Shahin}, ResNet and Inception nets were used for ensembling. Recently contrastive learning based approaches were explored in the FSL setting for skin lesion analysis~\cite{Yan2023}, and they show only marginal improvements.
However, as mentioned earlier, these approaches necessitate a large amount of labeled data, which is challenging to obtain.

\subsection{Semi-Supervised Learning based approaches}
SSL approaches gained popularity in skin lesion analysis~\cite{peng2023faxmatch,zhou2023refixmatch,zhou2023fixmatch,Mahmood2024,liu2020semi} as they harness both a small amount of labeled and a large amount of unlabeled data to enhance classification performance. By doing so, they decrease reliance on labeled data, offering a more efficient and effective approach to model training. 

FixMatch~\cite{fixmatch} is a prominent SSL methodology extensively adopted in the computer vision and deep learning communities. It merges principles from \emph{pseudo-labeling} and \emph{consistency regularization}. Pseudo-labeling involves assigning labels to unlabeled images during the training process based on the model's predictions. FixMatch employs both weak and strong augmentations. Initially, pseudo-labels for unlabeled images are inferred from weakly augmented versions. Subsequently, highly confident pseudo-labels are considered as genuine labels for strongly augmented images, facilitating model updates. This process enforces consistency between predictions made on weak and strong augmentations of the same unlabeled sample, thereby fostering model robustness. 

Building upon the achievements of FixMatch, several variants such as FixMatch-LS~\cite{zhou2023fixmatch}, ReFixMatch?LS~\cite{zhou2023refixmatch}, and FaxMatch~\cite{peng2023faxmatch} have emerged specifically for skin lesion classification. These methodologies incorporate label smoothing techniques to counteract overfitting and to generate trustworthy pseudo-labels. Moreover, they integrate diverse terms into the loss function, demonstrating improved performance compared to FixMatch. For example, FixMatch-LS~\cite{zhou2023fixmatch} introduces a sample relation consistency loss, aligning the predictions of teacher and student models. The Relation-driven Self-Ensembling model, as introduced in~\cite{liu2020semi}, ensures consistency between the outputs of the teacher and student models under various perturbations of the same input. Conversely, the recently introduced SPLAL~\cite{Mahmood2024} leverages the outputs of multiple classifiers to determine pseudo-labels, thereby improving the reliability of the selected pseudo-labels.

In contrast to these methods, our proposed approach leverages ensemble learning and online knowledge distillation to achieve superior performance.

\section{Methodology\label{sec:method}}
\begin{figure}[!t]
	\centering
	\includegraphics[width=\textwidth]{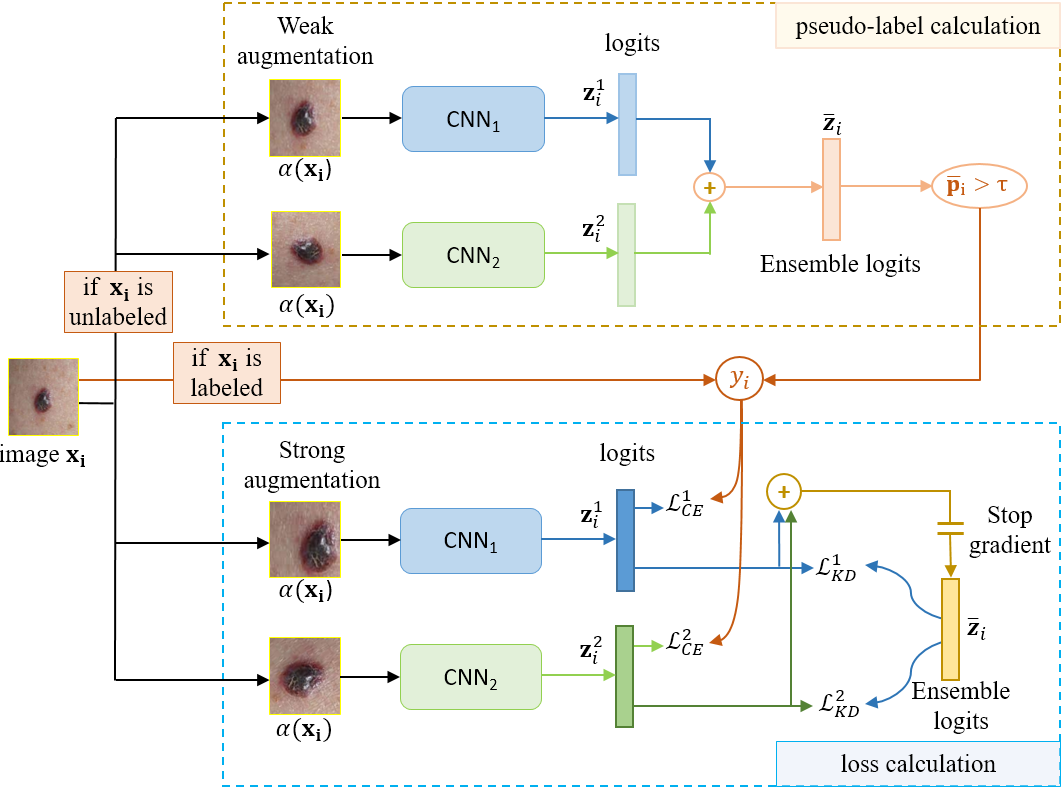}
	\caption{Overview of the proposed approach: Utilizing an ensemble size $K=2$, weakly augmented unlabeled images are used to generate ensemble predictions, from which pseudo-labels are derived. Subsequently, strong augmentations are employed during loss calculations. Each ensemble member undergoes training with a combination of two loss functions: one optimizing classification accuracy via cross-entropy loss ($\mathcal{L}^k_{CE}$), while the other facilitates knowledge distillation from the ensemble model to individual members via distillation loss ($\mathcal{L}^k_{KD}$). To enhance computational and memory efficiency, gradient updates  are not backpropagated (stop gradient) from ensemble logits. \label{fig:overview}}
\end{figure}

This work introduces a novel \emph{self-training}-based~\cite{amini2023selftraining} semi-supervised learning (SSL) framework for skin lesion classification. As shown in Figure~\ref{fig:overview} and detailed in Algorithm~\ref{alg}, our approach leverages both a limited set of labeled images and a large pool of unlabeled images to enhance classification performance.

Assume that the training data consists of a small labeled set and a large unlabeled set. Let
$\mathcal{D}_L = \{ (\mathbf{x}_i, y_i) \}$
represent the labeled dataset, where each image \(\mathbf{x}_i\) is paired with a label \(y_i \in \{1, \dots, C\}\) from \(C\) classes. The unlabeled dataset is given by
$\mathcal{D}_U = \{ \mathbf{x}_i \},$
comprising images without known labels.

Our framework consists of two stages:

\begin{enumerate}
	\item \textbf{Supervised Training with Knowledge Distillation:} In the first stage, an ensemble of \(K\) CNN models is trained using \(\mathcal{D}_L\). Each model is optimized with a composite loss that combines cross-entropy loss which maximizes classification accuracy, and a knowledge distillation loss that transfers the ensemble's collective knowledge to individual models.
	
	\item \textbf{Pseudo-Labeling and Dataset Expansion:} In the second stage, the ensemble is used to assign pseudo-labels to unlabeled samples from \(\mathcal{D}_U\). High-confidence pseudo-labels (i.e., those exceeding a predefined threshold) are added to the labeled dataset, and the models are retrained on this expanded set. This pseudo-labeling and retraining process is iterated (e.g., three times) to progressively improve the quantity of labeled data.
\end{enumerate}

In addition, following~\cite{fixmatch}, we apply both weak and strong data augmentations. Weak augmentation (e.g., center cropping) is used for generating pseudo-labels, whereas strong augmentation (e.g., random cropping) is applied during training to update the model weights.

\newcommand\mycommfont[1]{\footnotesize\ttfamily\textcolor{blue}{#1}}
\SetCommentSty{mycommfont}
\begin{algorithm}[!t]
	\small 
	\DontPrintSemicolon
	\KwInput{Labeled set $\mathcal{D}_L$, unlabeled set $\mathcal{D}_{U}$, $K$ - size of the ensemble, $n_{epoch}$ -- number of epochs, $n_{iter}$ -- number of SSL iterations.}
	\KwOutput{CNN ensemble $f_e = \{f^1,\dots,f^K\}$}
	\KwData{}
	\For{t=1 to $n_{iter}$}    
	{ 
		\tcc{Supervised Training with Knowledge Distillation}
		\For{epoch=1 to $n_{epoch}$} 
		{
			\tcc{Calculation of soft-probabilities for knowledge distillation}
			\For{k=1 to K} 
			{
				calculate logits $\mathbf{z}^k = f^k(\alpha(\mathbf{x}))$
			}
			calculate $\bar{\mathbf{p}}_s$  using Eqn.~\ref{eqn:softprob}\\
			\vspace{5pt}
			\tcc{Train each model of the ensemble}
			\For{k=1 to K} 
			{
				optimize the weights of the $k^\text{th}$ CNN model using~Eqn~\ref{eqn:floss}\\
		}}
		\vspace{5pt}
		\tcc{Augment the labeled  set}		
		generate pseudo-labeles for the weakly augmented $\mathcal{D}_U$ using ensemble (Sec.~\ref{sec:pl})\\
		form $\mathcal{D}'_L$ by selecting high confident pseudo-labels from $\mathcal{D}_U$ (Sec.~\ref{sec:pl})\\
		update the labeled and unlabeled sets: \\ \Indp$\mathcal{D}_L = \mathcal{D}_L \cup \mathcal{D}'_L$\\
		$\mathcal{D}_U = \mathcal{D}_U \setminus \mathcal{D}'_L$\\
		\Indm
	}
	\caption{Ensemble based semi-supervised deep learning with online knowledge distillation.\label{alg}}
\end{algorithm}

\subsection{Supervised Training with Knowledge Distillation\label{sec:loss}}
Each of the \(K\) CNN models is trained on the labeled dataset
$\mathcal{D}_L$. A stochastic augmentation function \(\alpha(\cdot)\) is applied to each image before feeding it to the network.

The overall loss for the \(k^\text{th}\) model, \(\mathcal{L}^k\), is a weighted combination of the cross-entropy loss \(\mathcal{L}_{CE}^k\) and the knowledge distillation loss \(\mathcal{L}_{KD}^k\):
\begin{equation}
	\mathcal{L}^k = \mathcal{L}_{CE}^k + \lambda \, \mathcal{L}_{KD}^k,
	\label{eqn:floss}
\end{equation}
where \(\lambda\) is a trade-off parameter that balances the two loss components.

\subsubsection{Cross-Entropy Loss}
We adopt the standard cross-entropy (CE) loss for multi-class classification to maximize correct predictions. For model \(f^k\) with logits \(\mathbf{z}_i^k\) obtained for an augmented image \(\alpha(\mathbf{x}_i)\), the softmax function yields the predicted probability for class \(c\):
\begin{equation}
	p(c|\alpha(\mathbf{x}_i), f^k) = \frac{\exp\left(z_{ic}^k\right)}{\sum_{j=1}^{C} \exp\left(z_{ij}^k\right)}.
\end{equation}

Due to class imbalance in skin lesion datasets, we incorporate class weights \(w_c\) (computed via normalized inverse class frequencies) to ensure fair contribution of each class. The CE loss for the \(k^\text{th}\) model over \(N\) labeled samples is given by:
\begin{equation}
	\mathcal{L}_{CE}^k = -\frac{1}{N}\sum_{c=1}^{C}w_c\sum_{i=1}^{N} \mathbbm{1}_{\{y_i=c\}} \log p(c|\alpha(\mathbf{x}_i), f^k),
	\label{eqn:ce}
\end{equation}
where \(\mathbbm{1}_{\{y_i=c\}}\) is an indicator function that equals 1 when \(y_i = c\) and 0 otherwise, and the class weights are computed as:
\begin{equation}
	w_c = \frac{\frac{1}{n_c}}{\sum_{i=1}^{C}\frac{1}{n_i}},
	\label{eqn:cw}
\end{equation}
with \(n_c\) representing the number of samples in class \(c\).

\subsubsection{Knowledge Distillation Loss}
While ensemble models typically yield superior performance~\cite{ganaie2022ensemble}, deploying them during testing can be resource-intensive. To address this, we use online knowledge distillation to transfer the ensemble's knowledge to each individual model, enabling any single model to perform comparably to the ensemble.

For each model \(f^k\), we first compute the \emph{softened} class probabilities with a temperature parameter \(T\)~\cite{hinton2015distilling}:
\begin{equation}
	p_s(c|\alpha(\mathbf{x}_i), f^k) = \frac{\exp\left(z_{ic}^k/T\right)}{\sum_{j=1}^{C} \exp\left(z_{ij}^k/T\right)}.
	\label{eqn:softprobk}
\end{equation}

The ensemble's prediction is obtained by averaging the logits from all \(K\) models:
\begin{equation}
	\bar{\mathbf{z}}_i = \frac{1}{K}\sum_{k=1}^{K} \mathbf{z}_i^k.
	\label{eqn:z}
\end{equation}
The softened ensemble probabilities are then defined as:
\begin{equation}
	\bar{p}_s(c|\alpha(\mathbf{x}_i), f^e) = \frac{\exp\left(\bar{z}_{ic}/T\right)}{\sum_{j=1}^{C} \exp\left(\bar{z}_{ij}/T\right)}.
	\label{eqn:softprob}
\end{equation}

The knowledge distillation loss for the \(k^\text{th}\) model is formulated to align its softened predictions with those of the ensemble. Averaged over \(N\) samples, the loss is:
\begin{equation}
	\mathcal{L}_{KD}^k = -\frac{T^2}{N} \sum_{i=1}^{N} \sum_{c=1}^{C} \bar{p}_s(c|\alpha(\mathbf{x}_i), f^e) \log p_s(c|\alpha(\mathbf{x}_i), f^k).
	\label{eqn:kd}
\end{equation}
The factor \(T^2\) balances the gradients when using a higher temperature.

\subsection{Pseudo-Labeling and Dataset Expansion\label{sec:pl}}
To further leverage unlabeled data, we employ a pseudo-labeling strategy. The ensemble's unsmoothed prediction for an image \(\alpha(\mathbf{x}_i)\) is given by:
\begin{equation}
	\bar{p}(c|\alpha(\mathbf{x}_i), f^e) = \frac{\exp\left(\bar{z}_{ic}\right)}{\sum_{j=1}^{C} \exp\left(\bar{z}_{ij}\right)},
\end{equation}
and the pseudo-label \(\hat{y}_i\) is determined by:
\begin{equation}
	\hat{y}_i = \arg\max_{c} \; \bar{p}(c|\alpha(\mathbf{x}_i), f^e), \quad c = 1,\dots, C.
\end{equation}

Only high-confidence pseudo-labels, where the maximum probability exceeds a threshold \(\tau\) (i.e., \(\max_{c} \bar{p}(c|\alpha(\mathbf{x}_i), f^e) > \tau\)), are selected. These high-confidence samples, along with their pseudo-labels, are then incorporated into the original labeled set \(\mathcal{D}_L\). The expanded dataset is used to retrain the models, and this supervised training and pseudo-labeling cycle is repeated for a fixed number of iterations (e.g., 3 iterations) to progressively expand the labeled training set.

\noindent In summary, our methodology combines supervised training with knowledge distillation and iterative pseudo-labeling to harness both labeled and unlabeled data for robust skin lesion classification.

\section{Experiments, Results and Discussion\label{sec:exp}}
\subsection{Datasets and Implementation Details}
\subsubsection{Datasets}
This study evaluated the proposed approach using two widely recognized skin lesion classification datasets, ISIC 2018 and ISIC 2019. These datasets, extensively used in related literature (e.g., \cite{zhou2023refixmatch, zhou2023fixmatch, liu2020semi,Mahmood2024}), consist of 10,015 and 25,331 images across 7 and 8 classes, respectively. The class distribution in these datasets is notably imbalanced, as shown in Table~\ref{tab:datasets}.

\begin{table}[t!]
	\centering
	\renewcommand{\arraystretch}{1.5}
	\renewcommand{\tabcolsep}{3pt}
	\begin{center}
		\caption{Statistics of the datasets}
		\label{tab:datasets}
		{\scriptsize
			\begin{tabular}{|l|R{1.6cm}|R{1.6cm}|}\hline
				\textbf{Class Name} & \textbf{ISIC 2018} & \textbf{ISIC 2019} \\ \hline
				Melanoma & 1,103& 4,522\\ \hline
				Melanocytic nevus& 6,716 &12,875 \\ \hline
				Basal cell carcinoma& 529 & 3,323\\ \hline
				Actinic keratosis& 325 & 867\\ \hline
				Benign keratosis&1,087 & 2,624\\ \hline
				Dermatofibroma& 120 &239 \\ \hline
				Vascular lesion&135 & 253\\ \hline
				Squamous cell carcinoma & -- & 628\\ \hline
				\textbf{Total} &\textbf{10,015} & \textbf{25,331}\\ \hline
			\end{tabular}
		}
	\end{center}
\end{table}

\subsubsection{Image Preprocessing and data augmentations\label{sec:preprocess}}
To address varying sizes and aspect ratios, images were initially resized to a height of $280$ pixels while maintaining their aspect ratio. Weak augmentation involved cropping a $224\times224$ pixel region from the center, followed by random rotations, and random horizontal and vertical flipping. For strong augmentation, a random $224\times224$ pixel crop was used, instead of center crop. During testing, no augmentation was applied, except extracting a $224\times224$ pixel region from the center of the resized image. The preprocessed images were normalized to have zero mean and unit variance.

\subsubsection{Evaluation Criteria}
To address the challenge of imbalanced data, we employed \emph{balanced accuracy} (BAcc), specifically the average of per-class recall values as follows:
\begin{equation}
	{BAcc} = \frac{1}{C}\sum_c \frac{TP_c}{TP_c+FN_c}
\end{equation} 
where $TP_c$ and $FP_c$ represents the number of true positives and the false negatives respectively) obtained on each class, as the primary evaluation metric. Additionally, we considered overall accuracy to gauge the general prediction performance, which represents the percentage of the overall correct predictions. In addition, we reported the macro-F1 score, a widely adopted metric in pertinent literature~\cite{zhou2023refixmatch, zhou2023fixmatch, liu2020semi}, for method comparison. The macro-F1 score is the average of per-class F1-scores as given below:
\begin{equation}
	F1 = \frac{1}{C}\sum_c \frac{2 TP_c}{2 TP_c + FP_c + FN_c}
\end{equation}
where, $FP_c$ and $FN_c$ respectively represent the number of false positive and false negatives  calculated for the class $c$.

For a comprehensive assessment, we compared the accuracy values as reported in~\cite{zhou2023refixmatch, zhou2023fixmatch, liu2020semi}, denoted as Acc* as follows:
\begin{equation}
	Acc* = \frac{1}{C}\sum_c Acc_c
\end{equation} 
Here, $Acc_c$ is the binary class accuracy, calculated by considering the class $c$ as the positive class and all the other classes as the negative class.
It is important to note that Acc* amplifies the challenges associated with imbalanced classification, exhibiting a notable bias towards the correct prediction of the majority class. For instance, consider a three-class classification problem with 80, 10, and 10 samples in each class. If a classifier predicts only the samples from the majority class correctly, the Acc* score would still be high at 80\%. However, the BAcc in this scenario would be only 27\%, accurately reflecting the classifier's poor performance across the minority classes. Therefore, we argue that although Acc* has been used by the existing literature~\cite{zhou2023refixmatch, zhou2023fixmatch, liu2020semi} to compare different methods, relying on Acc* for performance comparison is highly ineffective.

\subsubsection{Experimental Settings}
The proposed method employed an ImageNet pretrained ResNet-50~\cite{HeZRS16} as the backbone network. We utilized the Adam optimizer coupled with cosine learning rate decay, where the learning rate at epoch $m$ was determined as $\eta \cos(\frac{7\pi m}{18M})$, with $M$ representing the total number of epochs set to 40. Batch size was set to 32. To mitigate overfitting, a dropout rate of 0.5 was applied at the layer preceding the classification layer. 

In accordance with \cite{zhou2023refixmatch, zhou2023fixmatch, liu2020semi}, we partitioned images from each class in both datasets into training (70\%), validation (10\%), and testing (20\%) sets. The predefined splits for the ISIC 2018 dataset, as provided in \cite{liu2020semi}, were adopted to ensure a fair comparison. For the ISIC 2019 dataset, we generated the splits. All experiments were conducted thrice, and the reported results include the mean and standard errors of the evaluation measures. For all SSL experiments, a percentage $p\%$ of the training data was randomly selected as the labeled set, with the labels of the remaining data discarded for use as the unlabeled set. To ensure a fair comparison with previous studies such as \cite{zhou2023refixmatch, zhou2023fixmatch, liu2020semi}, we varied the value of $p$ across several intervals: $5\%$, $10\%$, $20\%$, and $100\%$ for the ISIC 2018 dataset, and $2\%$, $5\%$, $10\%$, and $100\%$ for the ISIC 2019 dataset, respectively. The threshold $\tau$ was set to $\tau=0.95$. Refer Section~\ref{sec:effect} for the effect of the value of $\tau$ in classification performance.

In the results, \textbf{Baseline} indicates the FSL approach trained with $p\%$ of the labeled training data. \textbf{Proposed ($K=k'$)} denotes the proposed SSL approach trained with $p\%$ of the labeled data, alongside the remaining unlabeled training data, where $k'$ represents the number of models in the ensemble. Additionally, \textbf{Proposed (KD)} represents the knowledge-distilled CNN model within the ensemble, reflecting the average results obtained across all models in the ensemble. It's important to note that \textbf{Proposed (KD)} is a single model, thus reducing computational complexities at test time compared to the ensemble model, while aiming to deliver superior performance compared to a model trained independently (i.e., \textbf{Proposed (K=1)}).

\subsection{Performance of the Proposed Approach}
Tables~\ref{tab:re2019} and \ref{tab:re2018} summarize the results of comparative experiments conducted on the ISIC 2019 and ISIC 2018 datasets. These results show: (1)~Increasing the quantity of labeled training data enhances overall performance. (2) Ensemble learning yields improved performance for both FSL and SSL settings compared to individual models. (3) SSL gives improved performance over FSL. (4) Employing distillation techniques enhances the performance of individual models.

\begin{table}[!t]
	\renewcommand{\arraystretch}{1.2}
	\renewcommand{\tabcolsep}{3pt}
	\begin{center}
		\caption{Performance of the proposed approach on the \textbf{ISIC 2019} dataset. The colors {\color{red}red}, {\color{blue}blue}, and {\color{cyan}cyan}  represent the rankings from first to third-best results based on their corresponding mean values.}
		\label{tab:re2019}
		{\scriptsize
			\begin{tabular}{cclccL{1.9cm}L{1.9cm}}\hline
				\multirow{2}{*}{p} & \multirow{2}{*}{Type} &\multirow{2}{*}{Method} &\multicolumn{4}{c}{Metrics} \\ \cmidrule{4-7}
				& & &BAcc	& Acc	&Acc*	& F1 \\ \Xhline{3\arrayrulewidth}
				\multirow{8}{*}{$2\%$ } & \multirow{2}{*}{FS}&\textbf{Ours:} Baseline (K = 1) &  $38.87 \pm 0.82$ &$62.85 \pm 0.28$ &$89.99 \pm 0.09$ &$39.24 \pm 0.96$ \\ \cmidrule{3-7} 
				&&\textbf{Ours:} Baseline (K = 3) &        $40.12 \pm 0.39$ &$\color{cyan}64.27 \pm 0.40$ &$\color{cyan}90.34 \pm 0.07$ &$40.62 \pm 0.55$\\ \cmidrule{2-7}
				
				&\multirow{6}{*}{SS} &FixMatch~\cite{fixmatch}$^\dag$ &     --   &   --   &    $  86.13\pm0.36 $  &     $ 38.37 \pm0.25  $  \\  \cmidrule{3-7}
				&&FixMatch-LS~\cite{zhou2023fixmatch} &   --   &   --   &    $ 87.72 \pm 0.08$  &     $ \color{cyan}42.74 \pm 0.24 $  \\  \cmidrule{3-7} 
				&&ReFixMatch-LS~\cite{zhou2023refixmatch} &     --   &   --   &    $87.76 $  &     $ 42.15 $  \\ \cmidrule{3-7} 
				&&\textbf{Ours:} Proposed (K = 1) & $\color{cyan}43.84 \pm 1.63$ &$63.64 \pm 0.29$ &$90.19 \pm 0.03$ &$42.15 \pm 1.51$\\ \cmidrule{3-7}
				&&\textbf{Ours:} Proposed (KD) & $\color{blue}45.07 \pm 0.83$ &$\color{blue}64.95 \pm 1.03$ &$\color{blue}90.51 \pm 0.24$ &$\color{blue}44.21 \pm 0.87$  \\ \cmidrule{3-7} 
				&&\textbf{Ours:} Proposed (K = 3) &  ${\color{red}\mathbf{45.39 \pm 0.81}}$ &${\color{red}\mathbf{65.30 \pm 1.10}}$ &${\color{red}\mathbf{90.60 \pm 0.27}}$ &${\color{red}\mathbf{44.60 \pm 0.85}}$ \\ \Xhline{3\arrayrulewidth} 

				\multirow{8}{*}{$5\%$ } & \multirow{2}{*}{FS}&\textbf{Ours:} Baseline (K = 1) &$46.72 \pm 0.46$ &$66.54 \pm 0.59$ &$90.92 \pm 0.15$ &$46.25 \pm 0.61$ \\ \cmidrule{3-7}
				&&\textbf{Ours:} Baseline (K = 3) &        $46.81 \pm 0.20$ &$\color{cyan}68.62 \pm 0.58$ &$\color{cyan}91.49 \pm 0.17$ &$47.33 \pm 0.64$\\ \cmidrule{2-7}
				&\multirow{6}{*}{SS} &FixMatch~\cite{fixmatch}$^\dag$ &     --   &   --   &    $  89.27\pm0.09 $  &     $ 46.25 \pm0.15  $  \\  \cmidrule{3-7} 
				&&FixMatch-LS~\cite{zhou2023fixmatch} &   --   &   --   &    $ 89.43 \pm 0.05$  &     $ 48.00 \pm 0.16 $  \\  \cmidrule{3-7} 
				&&ReFixMatch-LS~\cite{zhou2023refixmatch} &     --   &   --   &    $89.34 $  &     $ 46.70 $  \\ \cmidrule{3-7} 
				&&\textbf{Ours:} Proposed (K = 1) & $\color{cyan}49.64 \pm 0.52$ &$68.28 \pm 0.32$ &$91.38 \pm 0.11$ &$\color{cyan}48.51 \pm 0.34$ \\ \cmidrule{3-7}
				&&\textbf{Ours:} Proposed (KD) &    $\color{blue}50.88 \pm 0.65$ &$\color{blue}69.73 \pm 0.24$ &$\color{blue}91.76 \pm 0.07$ &$\color{blue}49.91 \pm 0.48$\\ \cmidrule{3-7} 
				&&\textbf{Ours:} Proposed (K = 3) &    ${\color{red}\mathbf{51.55 \pm 0.71}}$ &${\color{red}\mathbf{70.20 \pm 0.18}}$ &${\color{red}\mathbf{91.89 \pm 0.06}}$ &${\color{red}\mathbf{50.67 \pm 0.50}}$ \\ \Xhline{3\arrayrulewidth} 
				
				\multirow{8}{*}{$10\%$ } & \multirow{2}{*}{FS}&\textbf{Ours:} Baseline (K = 1) &    $52.56 \pm 0.83$ &$70.51 \pm 0.58$ &$92.00 \pm 0.17$ &$52.65 \pm 0.28$ \\ \cmidrule{3-7} 
				&&\textbf{Ours:} Baseline (K = 3) & $54.41 \pm 0.73$ &$\color{cyan}72.54 \pm 0.36$ &$92.52 \pm 0.12$ &$55.29 \pm 0.46$ \\ \cmidrule{2-7}
				&\multirow{6}{*}{SS} &FixMatch~\cite{fixmatch}$^\dag$ &     --   &   --   &    $  90.15\pm0.15 $  &     $ 50.71 \pm0.28  $  \\  \cmidrule{3-7} 
				&&FixMatch-LS~\cite{zhou2023fixmatch} &   --   &   --   &    $ 90.14 \pm 0.05$  &     $ 53.95 \pm 0.39 $  \\  \cmidrule{3-7} 
				&&ReFixMatch-LS~\cite{zhou2023refixmatch} &     --   &   --   &    $90.20 $  &     $ 54.60 $  \\ \cmidrule{3-7}
				&&\textbf{Ours:} Proposed (K = 1) & $\color{cyan}56.30 \pm 0.46$ &$72.47 \pm 0.28$ &$\color{cyan}92.55 \pm 0.09$ &$\color{cyan}56.02 \pm 0.31$ \\ \cmidrule{3-7}
				&&\textbf{Ours:} Proposed (KD) &     $\color{blue}56.42 \pm 0.90$ &$\color{blue}74.04 \pm 0.35$ &$\color{blue}92.93 \pm 0.08$ &$\color{blue}57.36 \pm 0.95$  \\ \cmidrule{3-7} 
				&&\textbf{Ours:} Proposed (K = 3) &     ${\color{red}\mathbf{56.63 \pm 0.94}}$ &${\color{red}\mathbf{74.38 \pm 0.36}}$ &${\color{red}\mathbf{93.01 \pm 0.08}}$ &${\color{red}\mathbf{57.80 \pm 0.99}}$  \\ \Xhline{3\arrayrulewidth} 
				
				\multirow{3}{*}{$100\%$ } & \multirow{2}{*}{FS }&\textbf{Ours:} Upper bound (K = 1)  &     $\color{blue}82.15 \pm 0.20$ &$86.00 \pm 0.09$ &$\color{blue}96.18 \pm 0.03$ &$\color{blue}81.49 \pm 0.30$ \\ \cmidrule{3-7} 
				&&\textbf{Ours:} Upper bound (K = 3)  &     ${\color{red}\mathbf{84.38 \pm 0.29}}$ &${\color{red}\mathbf{88.45 \pm 0.14}}$ &${\color{red}\mathbf{96.84 \pm 0.04}}$ &${\color{red}\mathbf{85.23 \pm 0.33}}$ \\
				\Xhline{3\arrayrulewidth}	
		\end{tabular}}
	\end{center}
	\end{table}
	
	\begin{table}[t!]
	\renewcommand{\arraystretch}{1.2}
	\renewcommand{\tabcolsep}{3pt}
	\begin{center}
		\caption{Performance of the proposed approach on the \textbf{ISIC 2018} dataset.}
		\label{tab:re2018}
		{\scriptsize
			\begin{tabular}{cclccL{1.9cm}L{1.9cm}}\hline
				\multirow{2}{*}{p} & \multirow{2}{*}{Type} &\multirow{2}{*}{Method} &\multicolumn{4}{c}{Metrics} \\ \cmidrule{4-7}
				& & &BAcc	& Acc	&Acc*	& F1 \\ \Xhline{3\arrayrulewidth}
				\multirow{8}{*}{$5\%$ } & \multirow{2}{*}{FS}&\textbf{Ours:} Baseline (K = 1) &$52.96 \pm 0.18$ &$75.03 \pm 0.12$ &$92.87 \pm 0.03$ &$53.68 \pm 0.54$ \\ \cmidrule{3-7}
				&&\textbf{Ours:} Baseline (K = 3) &$56.79 \pm 0.87$ &$\color{blue}76.29 \pm 0.43$ &$\color{blue}93.22 \pm 0.12$ &$57.04 \pm 0.92$ \\ \cmidrule{2-7}
				&\multirow{7}{*}{SS} &FixMatch~\cite{fixmatch}$^\dag$ &     --   &   --   &    $  89.70\pm0.30 $  &     $ 46.29 \pm0.31  $  \\  \cmidrule{3-7}
				&&FaxMatch~\cite{peng2023faxmatch} &   --   &   --   &    $ 91.20$  &     $ 52.65 $  \\  \cmidrule{3-7}
				
				&&FixMatch-LS~\cite{zhou2023fixmatch} &   --   &   --   &    $ 90.77 \pm 0.06$  &     $ 53.71 \pm 0.33 $  \\  \cmidrule{3-7}
				&&ReFixMatch-LS~\cite{zhou2023refixmatch} &     --   &   --   &    $90.74 $  &     $ 53.38 $  \\ \cmidrule{3-7}
				&&\textbf{Ours:} Proposed (K = 1) &     $\color{cyan}61.29 \pm 0.72$ &$74.51 \pm 0.62$ &$92.72 \pm 0.18$ &$\color{cyan}57.45 \pm 0.89$ \\ \cmidrule{3-7}
				&&\textbf{Ours:} Proposed (KD) &     $\color{blue}61.90 \pm 0.71$ &$\color{cyan}76.14 \pm 0.76$ &$\color{cyan}93.18 \pm 0.21$ &$\color{blue}58.84 \pm 0.33$ \\ \cmidrule{3-7}
				&&\textbf{Ours:} Proposed (K = 3) &  ${\color{red}\mathbf{62.62 \pm 0.54}}$ &${\color{red}\mathbf{76.63 \pm 0.68}}$ &${\color{red}\mathbf{93.32 \pm 0.19}}$ &${\color{red}\mathbf{59.99 \pm 0.28}}$ \\ \Xhline{3\arrayrulewidth}

				\multirow{8}{*}{$10\%$ } & \multirow{2}{*}{FS}&\textbf{Ours:} Baseline (K = 1) & $59.76 \pm 0.38$ &$77.23 \pm 0.38$ &$93.50 \pm 0.11$ &$58.02 \pm 0.47$ \\ \cmidrule{3-7}
				&&\textbf{Ours:} Baseline (K = 3) & $62.27 \pm 0.43$ &$\color{blue}79.20 \pm 0.50$ &$\color{blue}94.06 \pm 0.14$ &$\color{cyan}61.22 \pm 0.42$ \\ \cmidrule{2-7}
				&\multirow{6}{*}{SS} &FixMatch~\cite{fixmatch}$^\dag$ &     --   &   --   &    $  91.59\pm0.12 $  &     $ 54.20 \pm0.49  $  \\  \cmidrule{3-7}
				&&FaxMatch~\cite{peng2023faxmatch} &   --   &   --   &    $ 92.34$  &     $ 57.78 $  \\  \cmidrule{3-7}
				&&FixMatch-LS~\cite{zhou2023fixmatch} &   --   &   --   &    $ 91.93 \pm 0.16$  &     $ 59.40 \pm 0.32 $  \\  \cmidrule{3-7}
				&&ReFixMatch-LS~\cite{zhou2023refixmatch} &     --   &   --   &    $92.23 $  &     $ 59.36 $  \\ \cmidrule{3-7}
				&&\textbf{Ours:} Proposed (K = 1) & $\color{cyan}63.14 \pm 0.36$ &$77.44 \pm 0.50$ &$93.55 \pm 0.14$ &$59.78 \pm 0.13$ \\ \cmidrule{3-7}
				&&\textbf{Ours:} Proposed (KD) &   $\color{blue}64.70 \pm 0.89$ &$\color{cyan}78.84 \pm 0.62$ &$\color{cyan}93.95 \pm 0.18$ &$\color{blue}62.53 \pm 0.89$ \\ \cmidrule{3-7}
				&&\textbf{Ours:} Proposed (K = 3) &   ${\color{red}\mathbf{65.04 \pm 1.01}}$ &${\color{red}\mathbf{79.39 \pm 0.53}}$ &${\color{red}\mathbf{94.11 \pm 0.15}}$ &${\color{red}\mathbf{63.27 \pm 0.84}}$ \\ \Xhline{3\arrayrulewidth}

				\multirow{3}{*}{$100\%$ } & \multirow{2}{*}{FS }&\textbf{Ours:} Upper bound (K = 1)  &$\color{blue}75.02 \pm 0.41$ &$\color{blue}84.19 \pm 0.13$ &$\color{blue}95.48 \pm 0.04$ &$\color{blue}73.64 \pm 0.50$ \\ \cmidrule{3-7}
				&&\textbf{Ours:} Upper bound (K = 3)  &     ${\color{red}\mathbf{76.76 \pm 0.40}}$ &${\color{red}\mathbf{86.02 \pm 0.20}}$ &${\color{red}\mathbf{96.01 \pm 0.06}}$ &${\color{red}\mathbf{76.00 \pm 0.42}}$ \\
				\Xhline{3\arrayrulewidth}	
				
		\end{tabular}}
	\end{center}
\end{table}

The findings underscore the consistent enhancement in classification performance with increasing labeled training data, regardless of the learning methodology employed -- whether FSL or SSL. Notably, substantial improvements of approximately 40\% and 22\% in BAcc were observed on the ISIC 2019 and ISIC 2018 datasets, respectively, when transitioning from 5\% to 100\% labeled data using a single FSL-based model. These improvements persist even with ensemble learning.

Ensemble learning also proves to be beneficial, resulting in an approximate $2\%$ improvement in BAcc across various scenarios utilizing FSL. For instance, when only $5\%$ of labeled data is utilized with ISIC 2018 dataset, a single FSL based model achieves an BAcc of $52.96\pm 0.18$ and an F1 score of $53.68 \pm 0.54$. In contrast, an ensemble comprising three FSL models elevates these metrics to $56.79 \pm 0.87$ and $57.04 \pm 0.92$, respectively. It is important to note that as Acc* exhibits a notable bias towards the majority class, it yields values around $90\%$ even with a mere $5\%$ of labeled data. 

Diverse models are preferred to get improved performance by ensembling. Diverse models are favored to attain improved performance through ensembling. Although this study utilizes the ResNet-50 architecture initialized with ImageNet-trained weights, different random augmentations and dropout play a significant role in achieving model diversity during training. Incorporating differently initialized models is expected to yield even greater improvements through ensembling compared to a single model.

SSL yields substantial enhancements with both single and ensemble models. For instance, around $8\%$ improvements in BAcc was obtained with a single model when $5\%$ of labeled data is used with ISIC 2018 dataset. Considerable improvements were observed across other evaluation metrics, validating the effectiveness of SSL. Furthermore, when SSL is coupled with ensembling, the overall performance is further enhanced. For example, on the ISIC 2019 dataset with $5\%$ labeled data, SSL improved the F1 score by $2\%$ (from $46.25\%$ to $48.51\%$) compared to FSL. An additional $2\%$ improvement (to $50.67\%$) was achieved when combining ensembling, underscoring the effectiveness of both SSL and ensembling in enhancing classification performance, particularly in scenarios with limited labeled data.

Distillation improves the performance of individual models. For instance, when $10\%$ of labeled training data is used with ISIC 2018 dataset, an F1 score of $59.78\pm0.13$ was obtained by a single model. On the other hand, ensembling improves this performance to $63.27\pm0.84$. A knowledge distilled individual model achieves similar F1 score of $62.53\pm0.89$ proving the effectiveness of knowledge distillation to improve the classification performance of individual models, while reducing the computational complexity at test time.

\subsection{Comparison with the State-of-the-art}
This section establishes our proposed approach as the new state-of-the-art in skin lesion classification. Furthermore, it underscores the robustness of our FSL baselines.

\begin{table}[t!]
	\renewcommand{\arraystretch}{1.2}
	\renewcommand{\tabcolsep}{4.5pt}
	\begin{center}
		\caption{Comparison with the state-of-the-art methods on the \textbf{ISIC 2018} dataset with $20\%$ of the labeled data.}
		\label{tab:sota}
		{\scriptsize
			\begin{tabular}{clccL{1.9cm}L{1.9cm}}\hline
				\multirow{2}{*}{Type} &\multirow{2}{*}{Method} &\multicolumn{4}{c}{Metrics} \\ \cmidrule{3-6}
				& &BAcc	& Acc	&Acc*	& F1 \\ \Xhline{3\arrayrulewidth}

				\multirow{4}{*}{FS}&Baseline*~\cite{zhou2023refixmatch} &        --  &    --  &    $92.17$  &     $52.03$  \\  \cmidrule{2-6}
				&\textbf{Ours:} Baseline (K = 1) & $64.75 \pm 1.04$ &$79.63 \pm 0.54$ &$94.18 \pm 0.16$ &$62.40 \pm 0.71$ \\ \cmidrule{2-6}
				&\textbf{Ours:} Baseline (K = 3) & $68.39 \pm 0.67$ &$\color{cyan}81.55 \pm 0.18$ &$\color{cyan}94.73 \pm 0.05$ &$\color{cyan}66.71 \pm 1.10$ \\ \Xhline{3\arrayrulewidth}
				\multirow{22}{*}{SS}&Self-training~\cite{bai2017semi}   &     --  &    --& $92.37$ &$54.51$ \\ \cmidrule{2-6}
				&SS-DCGAN~\cite{diaz2019retinal} &       --  &    --& $92.27$ & $54.10$ \\ \cmidrule{2-6}
				&TCSE~\cite{li2018semi} &       --  &    --& $92.35$ & $58.44$ \\ \cmidrule{2-6}
				&TE~\cite{laine2016temporal} &       --  &    --& $92.26$ &$59.33$ \\ \cmidrule{2-6}
				&MT~\cite{tarvainen2017mean} &        --  &    --& $92.48$ &$59.10$ \\ \cmidrule{2-6}
				&SRC-MT~\cite{liu2020semi} &        --  &    --& $92.54$ &$60.68$ \\ \cmidrule{2-6}
				&FlexMatch~\cite{zhang2021flexmatch} &       --  &    --& $93.40\pm0.05$ &$61.37\pm0.02$ \\ \cmidrule{2-6}		
				&CoMatch~\cite{li2021comatch} &        --  &    --& $93.39\pm 0.04$ &$61.94\pm 0.21$ \\ \cmidrule{2-6}
				&FixMatch~\cite{fixmatch} &        --  &    --& $93.19\pm0.06$ &$60.12\pm0.23$ \\ \cmidrule{2-6}
				&{FaxMatch}~\cite{peng2023faxmatch} &        --  &    --& $93.19$ &$63.17$ \\ \cmidrule{2-6}
				&FixMatch-LS~\cite{zhou2023fixmatch} &        --  &    --& $93.25\pm0.05$ &$64.15\pm0.22$ \\ \cmidrule{2-6}
				&ReFixMatch-LS~\cite{zhou2023refixmatch} &        --  &    --& $93.40\pm0.04$ &$64.44\pm0.17$ \\ \cmidrule{2-6}
				&\textbf{Ours:} Proposed (K = 1) &     $\color{cyan}68.75 \pm 0.97$ &$80.25 \pm 0.39$ &$94.36 \pm 0.11$ &$65.21 \pm 0.64$ \\ \cmidrule{2-6}
				&\textbf{Ours:} Proposed (KD) &$\color{blue}69.39 \pm 0.82$ &$\color{blue}81.71 \pm 0.25$ &$\color{blue}94.87 \pm 0.08$ &$\color{blue}67.54 \pm 0.79$  \\     \cmidrule{2-6}
				&\textbf{Ours:} Proposed (K = 3) & ${\color{red}\mathbf{70.21 \pm 1.18}}$ &${\color{red}\mathbf{82.26 \pm 0.23}}$ &${\color{red}\mathbf{94.93 \pm 0.06}}$ &${\color{red}\mathbf{68.43 \pm 1.09}}$ \\  \hline
		\end{tabular}}
	\end{center}
\end{table}

Table~\ref{tab:sota} provides a comparison between our proposed approach and other state-of-the-art methods on the ISIC 2018 dataset, utilizing 20\% labeled training data. Additionally, Tables~\ref{tab:re2019} and ~\ref{tab:re2018} provide comparisons across both datasets with varying percentages of labeled training data. In all scenarios, our results outperform other approaches by a significant margin.

Notably, our FSL baseline demonstrates robust performance, achieving an F1 score of 62.40, surpassing the majority of existing FSL and SSL approaches in the literature, with the exception of FaxMatch, ReFixMatch-LS and FixMatch-LS models. Furthermore, our FSL-based ensemble model itself beats all the existing state-of-the-art FSL and SSL models, indicating the strength of our FSL models as robust baselines for comparison.

Our proposed SSL-based single model, ensemble model, and knowledge-distilled single model exhibit notable improvements in the F1 scores of approximately 1\%, 4\%, and 3\%, respectively, compared to the ReFixMatch-LS model -- the previous SSL state-of-the-art for skin lesion classification.

The results reveal that FixMatch and FlexMatch exhibit poor performance, largely attributed to their inability to effectively manage imbalanced class distributions. In contrast, our approach demonstrates proficiency in handling imbalanced data by appropriately weighting the contribution of different classes in the loss function.

\vspace{10pt}
\noindent\textit{Comparison with SPLAL~\cite{Mahmood2024}}: SPLAL, a recently introduced method, leverages class prototypes and a weighted combination of classifiers' outputs to enhance the accuracy of pseudo-label selection, reporting superior results on the ISIC 2018 dataset. Notably, SPLAL employs 80\% of the entire dataset for training, whereas our proposed approach utilizes 70\%, ensuring consistency with the majority of experiments reported in the literature.

\begin{table}[t!]
	\renewcommand{\arraystretch}{1.3}
	\renewcommand{\tabcolsep}{3pt}
	\begin{center}
		\caption{Comparison of our proposed approach with SPLAL~\cite{Mahmood2024} under the same experimental setting reported in~\cite{Mahmood2024}.}
		\label{tab:splal}
		{\scriptsize
			\begin{tabular}{cclccL{1.9cm}L{1.9cm}}\hline
				\multirow{2}{*}{p} & \multirow{2}{*}{Type} &\multirow{2}{*}{Method} &\multicolumn{4}{c}{Metrics} \\ \cmidrule{4-7}
				& & &BAcc	& Acc	&Acc*	& F1 score \\ \Xhline{3\arrayrulewidth}
				\multirow{6}{*}{$5\%$ } &\multirow{6}{*}{SS} &SPLAL~\cite{Mahmood2024} &     --   &   --   &    $  \color{red}\mathbf{93.58} $  &     $55.37$  \\  \cmidrule{3-7}
				&&\textbf{Ours:} Proposed (K=1) &    $\color{cyan}61.89 \pm 0.35$ &$\color{cyan}75.55 \pm 0.48$ &$93.01 \pm 0.14$ &$\color{cyan}57.39 \pm 1.01$  \\ \cmidrule{3-7}
				&&\textbf{Ours:} Proposed (KD) & $\color{blue}62.26 \pm 0.87$ &$\color{blue}76.26 \pm 0.23$ &$\color{cyan}93.22 \pm 0.07$ &$\color{blue}59.24 \pm 0.98$\\ \cmidrule{3-7}
				&&\textbf{Ours:} Proposed (K=3) &     ${\color{red}\mathbf{62.55 \pm 0.68}}$ &${\color{red}\mathbf{76.70 \pm 0.24}}$ &${\color{blue}93.34 \pm 0.07}$ &${\color{red}\mathbf{60.04 \pm 0.71}}$ \\ \Xhline{3\arrayrulewidth}

				\multirow{6}{*}{$20\%$ } &\multirow{6}{*}{SS} &SPLAL~\cite{Mahmood2024} &     --   &   --   &    $\color{blue}95.54$  &     $\color{cyan}70.46$  \\  \cmidrule{3-7}
				&&\textbf{Ours:} Proposed (K=1) &$\color{cyan}73.13 \pm 0.87$ &$\color{cyan}82.69 \pm 0.35$ &$95.05 \pm 0.10$ &$69.91 \pm 0.25$\\ \cmidrule{3-7}
				&&\textbf{Ours:} Proposed (KD) &$\color{blue}73.89 \pm 0.68$ &$\color{blue}84.07 \pm 0.23$ &$\color{cyan}95.45 \pm 0.06$ &$\color{blue}72.05 \pm 0.33$ \\ \cmidrule{3-7}
				&&\textbf{Ours:} Proposed (K=3) &${\color{red}\mathbf{74.56 \pm 0.57}}$ &${\color{red}\mathbf{84.62 \pm 0.12}}$ &${\color{red}\mathbf{95.61 \pm 0.03}}$ &${\color{red}\mathbf{73.06 \pm 0.47}}$ \\ \Xhline{3\arrayrulewidth}
		\end{tabular}}
	\end{center}
\end{table}

Table~\ref{tab:splal} presents the results of our approach under the same training setting as SPLAL. The findings demonstrate a significant advancement over SPLAL, validating the effectiveness of our proposed approach. For instance, while SPLAL reports an F1 score of 70.46, our proposed approach achieves an F1 score of $73.06$, representing approximately a 3\% improvement over SPLAL. This solidifies our proposed approach as the new state-of-the-art.

\subsection{Effect of parameters in the proposed approach\label{sec:effect}}
\begin{figure}[!h]
	\centering
	\begin{subfigure}{.5\textwidth}
		\centering
		\includegraphics[width=0.9\textwidth]{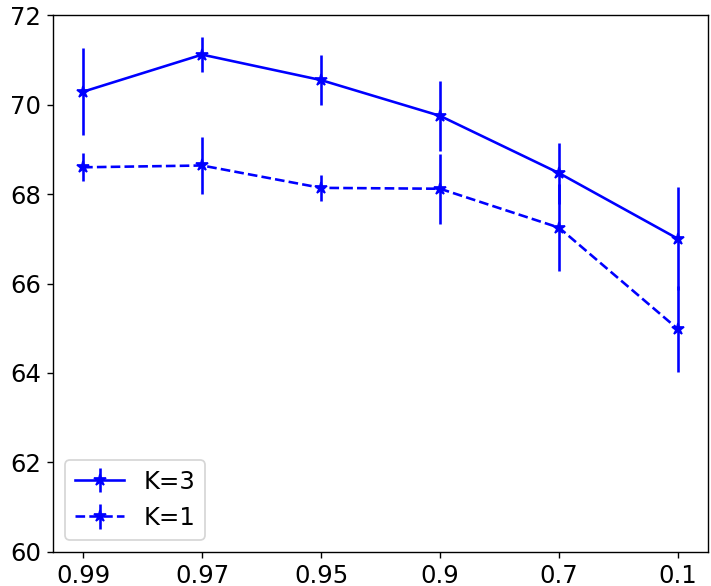}
		\caption{Threshold vs BAcc\label{fig:thr1}}
	\end{subfigure}%
	\begin{subfigure}{.5\textwidth}
		\centering
		\includegraphics[width=0.9\textwidth]{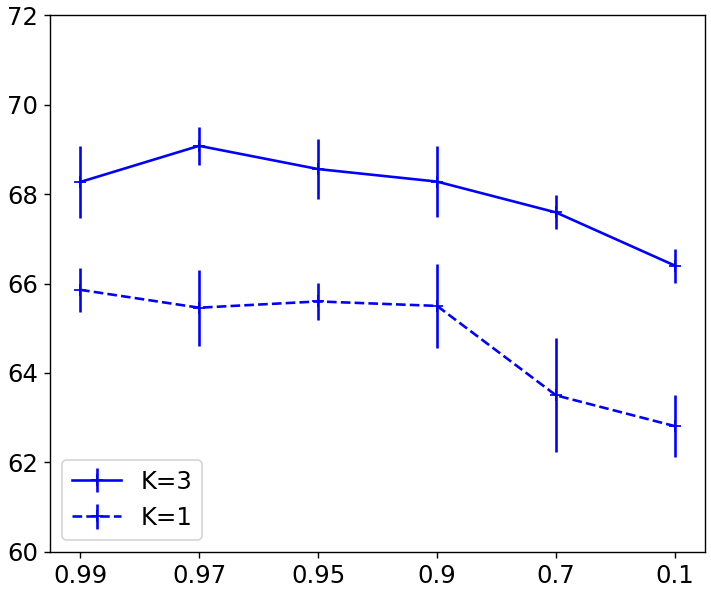}
		\caption{Threshold vs F1 score\label{fig:thr2}}
	\end{subfigure}
	\caption{Effect of threshold $\tau$ (x-axis) in the proposed SSL approach with $20\%$ of labeled training data}
\end{figure}

Figures~\ref{fig:thr1} and \ref{fig:thr2} depict the BAcc and F1 score, respectively, across varying thresholds using $20\%$ of labeled data for training on the ISIC 2018 dataset, with and without ensembling. These figures underscore the substantial performance improvement of the ensembled model over a single model. This also shows that better performance can be obtained by carefully choosing the threshold. A very high threshold ($\tau=0.99$) yields better performance with a single model, on the other hand, the ensemble model gives better results when threshold is $\tau=0.97$. Conversely, a low threshold ($\tau=0.1$) yields poor performance, attributed to the inclusion of numerous unreliable samples, indicative of confirmation bias or noise accumulation~\cite{Arazo}.

\begin{figure}[!h]
	\centering
	\begin{minipage}[t]{.47\textwidth}
		\centering
		\includegraphics[width=\textwidth, height=4.5cm]{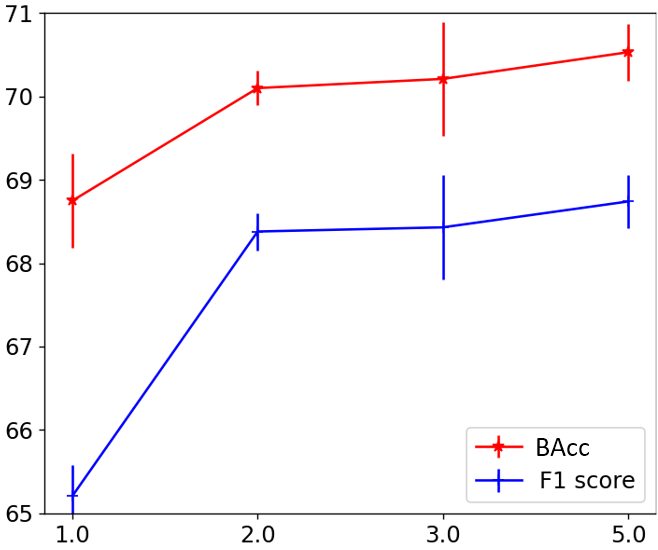}
		\caption{Ensemble size (x-axis) vs classification performance (y-axis).}
		\label{fig:effectEnSize}
	\end{minipage}\qquad %
	\begin{minipage}[t]{.47\textwidth}
		\centering
		\includegraphics[width=\textwidth, height=4.5cm]{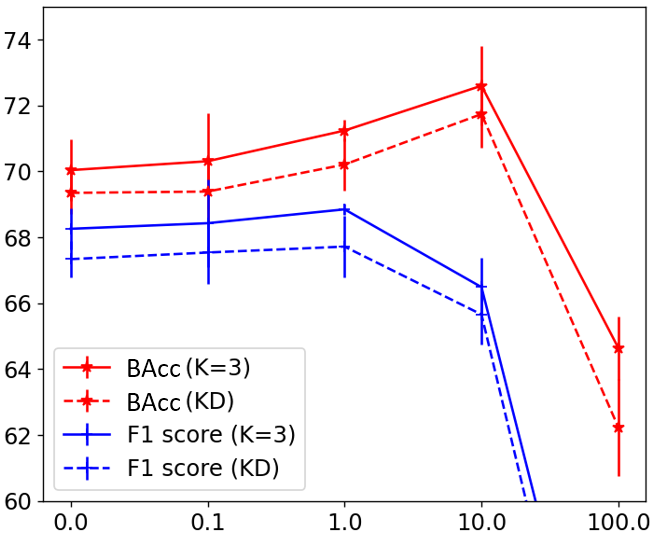}
		\caption{$\lambda$ (x-axis) vs classification performance (y-axis).}
		\label{fig:effectLambda}
	\end{minipage}
\end{figure}

Figure~\ref{fig:effectEnSize} demonstrates the impact of increasing the number of models in the ensemble on BAcc, Acc, and F1 score when utilizing $20\%$ of labeled training data from the ISIC 2018 dataset with SSL. The results reveal that increasing the ensemble size enhances overall performance. For instance, a notable improvement of approximately $2\%$ is observed in BAcc, Acc, and F1 score when transitioning from an ensemble size of $K=1$ to $K=3$. Subsequent increases in ensemble size yield only marginal performance gains compared to $K=3$.

Figure~\ref{fig:effectLambda} illustrates the impact of KD within the overall loss function. While the CE loss prioritizes enhancing BAcc, increasing the value of $\lambda$ enhances the classification performance of individual models, thereby improving the overall ensemble's performance. For instance, a $1\%$ enhancement in BAcc was observed when $\lambda$ was increased from 0 to 10. However, excessive emphasis on KD can diminish performance, as it detracts focus from the classification loss.

\subsection{Computational Complexity \label{sec:compu}}
All experiments were conducted on a single NVIDIA P100 GPU with 16GB of memory. When using 20\% of the labeled data, a fully supervised single model requires approximately 10 minutes to complete 40 epochs of training. This training time scales linearly with the number of models; for example, an ensemble with \(K=3\) models takes roughly 30 minutes.

In the semi-supervised setting, training times increase: a single model requires about 60 minutes, while an ensemble of 3 models takes approximately 180 minutes. Despite these differences in training time, inference remains efficient. Testing 1000 images takes around 3 seconds for a single model and 8 seconds for an ensemble with \(K=3\).

It is important to note that a single ResNet-50 model has approximately 26 million parameters, and the total parameter count scales linearly with the number of models, which can impose significant memory demands during testing. However, thanks to our knowledge distillation strategy, any individual model can be deployed at test time instead of the entire ensemble. This distilled model mimics the ensemble?s performance, reducing memory requirements while delivering improved performance compared to a standalone model.

\section{Conclusion\label{sec:con}}
This work proposed an ensemble-based semi-supervised deep learning method wherein multiple semi-supervised models are trained, and their collective knowledge is distilled into each model within the ensemble in an online fashion. This online distillation process enhances the performance of individual models within the ensemble, and consequently the performance of the ensemble. Following the training of the ensemble model using online distillation in a semi-supervised manner, any individual model from the ensemble can be deployed during the testing phase, reducing computational and memory demands. We showed that the knowledge distilled individual model performs better than the model which is trained independently. Our proposed approach reports new state-of-the-art results on both the ISIC 2018 and ISIC 2019 skin lesion analysis benchmark datasets, proving the efficiency of the proposed approach.

{\footnotesize
	\setlength{\bibsep}{0pt plus -0.1ex}
	\bibliography{references}
	\addcontentsline{toc}{section}{References}
}

\end{document}